\documentclass[runningheads]{llncs}
\usepackage[T1]{fontenc}
\usepackage{graphicx}
\usepackage{amsmath,amssymb}
\usepackage{xcolor}
\usepackage{array,multirow}
\newcolumntype{P}[1]{>{\centering\arraybackslash}p{#1}}
\newcolumntype{M}[1]{>{\centering\arraybackslash}m{#1}}
\usepackage{cite}
\begin{document}
\title{Diffusion Deformable Model for 4D Temporal Medical Image Generation}
\titlerunning{Diffusion Deformable Model for 4D Temporal Medical Image Generation}

\author{Boah Kim \orcidID{0000-0001-6178-9357} \and
Jong Chul Ye \orcidID{0000-0001-9763-9609}} 
\authorrunning{B. Kim and J. C. Ye}
%
\institute{Korea Advanced Institute of Science and Technology, Daejeon, South Korea \\
\email{\{boahkim, jong.ye\}@kaist.ac.kr}}
\maketitle            
\begin{abstract} 
Temporal volume images with 3D+t (4D) information are often used in medical imaging to statistically analyze temporal dynamics or capture disease progression. Although deep-learning-based generative models for natural images have been extensively studied, approaches for temporal medical image generation such as 4D cardiac volume data are limited. In this work, we present a novel deep learning model that generates intermediate temporal volumes between source and target volumes. Specifically, we propose a diffusion deformable model (DDM) by adapting the denoising diffusion probabilistic model that has recently been widely investigated for realistic image generation. Our proposed DDM is composed of the diffusion and the deformation modules so that DDM can learn spatial deformation information between the source and target volumes and provide a latent code for generating intermediate frames along a geodesic path. Once our model is trained, the latent code estimated from the diffusion module is simply interpolated and fed into the deformation module, which enables DDM to generate temporal frames along the continuous trajectory while preserving the topology of the source image. We demonstrate the proposed method with the 4D cardiac MR image generation between the diastolic and systolic phases for each subject. Compared to the existing deformation methods, our DDM achieves high performance on temporal volume generation.

\end{abstract}
%
%
\section{Introduction}
Exploring the progression of anatomical changes is one of the important tasks in medical imaging for disease diagnosis and therapy planning. In particular, for the case of cardiac imaging with inevitable motions, 4D imaging to monitor 3-dimensional (3D) volume changes according to the time is often required for accurate analysis \cite{amsalu2021spatial}.  Unfortunately, in contrast to CT and ultrasound, MRI requires a relatively long scan time to obtain the 4D images.

With the advances in deep learning approaches, deep learning-based medical image generation methods have been extensively developed \cite{nie2017medical, yang2018unpaired, nie2018medical, dai2020multimodal}. However, the existing models based on generative adversarial networks (GAN) may generate artificial features, which should not be occurred in the medical imaging area. On the other hand, learning-based deformable image registration methods have been developed due to their capability to provide deformation fields in real-time for the source to be warped into the target \cite{balakrishnan2018unsupervised, balakrishnan2019voxelmorph, dalca2018unsupervised, kim2021cyclemorph}. The smooth deformation fields enable the source to deform with topology preservation. Using this property, although the generative methods employing the registration model are presented \cite{dalca2019learning, dey2021generative}, the role of these models is to generate templates instead of generating temporal 3D volume frames along the continuous trajectory.

Recently, the denoising diffusion probabilistic model (DDPM) \cite{ho2020denoising, sohl2015deep} has been shown impressive performance in generating realistic images by learning a Markov chain process for the transformation of the simple Gaussian distribution into the data distribution \cite{choi2021ilvr, saharia2021image, song2020denoising}. Inspired by the fact that DDPM generates images through the latent space provided by the parameterized Gaussian process, the main contribution of this work is to present a novel diffusion deformable model (DDM) composed of the diffusion and deformation modules so that the latent code has detailed spatial information of the source image toward the target. Specifically, given the source and target images, the latent code is learned by using a score function of the DDPM, which is fed into the deformation module to generate temporal frames through the deformation fields.

Once the proposed DDM is trained, intermediate frame 4D temporal volumes can be easily generated by simply interpolating the latent code estimated by the diffusion module. More specifically, the deformation module of DDM estimates the deformation fields according to the scaled latent code and provides the continuous deformation from the source to the target images in the image registration manner, which can preserve the topology of the source image. In this sense, our score-based latent code can be interpreted to provide plausible geodesic paths between the source and target images.

We verify the proposed DDM on 4D cardiac MR image generation using ACDC dataset \cite{bernard2018deep}. The experimental results show that our model generates realistic deformed volumes along the trajectory between the diastolic and systolic phases, which outperforms the methods of adjusting the registration fields. The main contributions of this work can be summarized as follows:
\begin{itemize}
    \item We propose a diffusion deformable model for 4D medical image generation, which employs the denoising diffusion model to estimate the latent code. 
    \item By simply scaling the latent code, our model provides non-rigid continuous deformation of the source image toward the target. 
    \item Experimental results on 4D cardiac MRI verify that the proposed method generates realistic deformed images along the continuous trajectory.
\end{itemize}

\begin{figure}[t!]
\centering
\includegraphics[width=\textwidth]{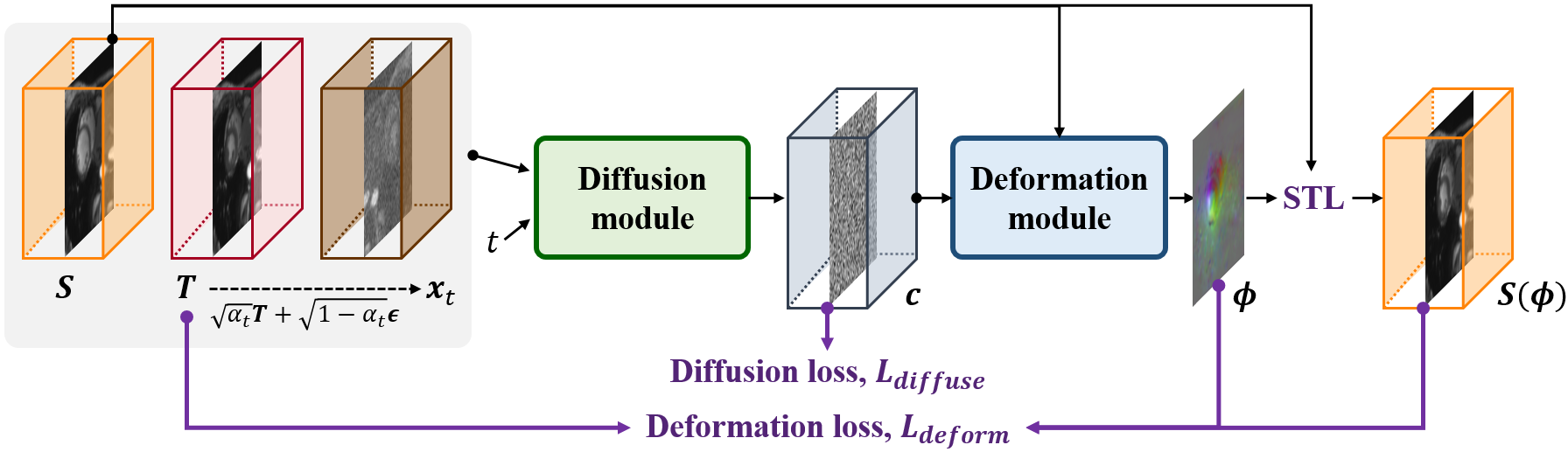}
\caption{The overall framework of the proposed method. The diffusion module takes a source $S$, a target $T$, and the perturbed target $x_t$, and estimates the latent code $c$. Then, the deformation module takes the code $c$ with the source $S$ and generates the deformed image $S(\phi)$ using the deformation fields $\phi$ with the spatial transformation layer (STL).} 
\label{fig:method}
\end{figure}

\section{Proposed Method}
The overall framework of the proposed diffusion deformable model (DDM) is illustrated in Fig.\ref{fig:method}. For the source $S$ and the target $T$ volumes, we design the model composed of a diffusion module and a deformation module, which is trained in an end-to-end manner. When the diffusion module estimates the latent code $c$ that provides the spatial information of intermediate frames between the source and target, the deformation module generates deformed images along the trajectory according to the code. More details are as follows.

\subsection{Diffusion Deformable Model}
\subsubsection{Diffusion Module}
The diffusion module is based on the denoising diffusion probabilistic model (DDPM) \cite{ho2020denoising, sohl2015deep}. DDPM is proposed as a generative model that learns a Markov chain process to convert the Gaussian distribution into the data distribution. The forward diffusion process adds noises to the data $x_0$, so that the data distribution at $t\in [0, u]$ of $x$ can be represented as:
\begin{equation}
    q(x_t|x_{t-1}) = \mathcal{N} (x_t; \sqrt{1-\beta_t}x_{t-1}, \beta_t \textbf{I}), 
\end{equation}
where $\beta_t$ is a noise variance in the range of (0, 1). Then, for the reverse deiffusion, DDPM learns the following parameterized Gaussian transformations:
\begin{equation}
    p_\theta (x_{t-1}|x_t)=\mathcal{N}(x_{t-1};\mu_\theta (x_t, t), \sigma_t^2 \textbf{I}),
\end{equation}
where $\mu_\theta (x_t, t)$ is learned mean and $\sigma_t$ is a fixed variance. Accordingly, the generative process to sample the data is performed by the stochastic step: $x_{t-1}=\mu_\theta (x_t, t)+\sqrt{\beta_t}z$, where $z \sim \mathcal{N}(0, \textbf{I})$.

By employing the property of DDPM, we design the diffusion module by considering the target as the reference data, i.e. $x_0=T$. Specifically, given the condition of source $S$ and target $T$, we first sample the perturbed target $x_t$ by:
\begin{equation}
    x_t = \sqrt{\alpha_t}T+\sqrt{1-\alpha_t}\epsilon,
\end{equation}
where $\epsilon \sim \mathcal{N}(0, \textbf{I})$, under $q(x_t|x_0) = \mathcal{N} (x_t; \sqrt{\alpha_t}x_0, (1-\alpha_t)\textbf{I})$ where $\alpha_t = \Pi_{s=1}^t (1-\beta_s)$. Then, using the condition and the perturbed target $x_t$, the diffusion module learns the latent code $c$ that compares the source and target and contains information of the score function for deformation.

\subsubsection{Deformation Module} 
When the diffusion module outputs the latent code $c$, the deformation module generates the deformed image according to the code $c$. This module leverages the learning-based image registration method \cite{balakrishnan2018unsupervised, balakrishnan2019voxelmorph}. Specifically, as shown in Fig.~\ref{fig:method}, the deformation module estimates the registration fields $\phi$ for the source image $S$. Then, using the spatial transformation layer (STL) \cite{jaderberg2015spatial} with tri-linear interpolation, the deformed source image $S(\phi)$ is generated by warping the volume $S$ with $\phi$. Here, unlike \cite{balakrishnan2018unsupervised, balakrishnan2019voxelmorph} that directly takes the source and target images as a network input, the deformation module of the proposed DDM takes the latent code $c$ and the source. As will be shown later, this enables our model to provide continuous deformation according to the code.

\subsection{Model Training}
To train the proposed DDM, we design the loss function as follows:
\begin{equation}
    \min_{G_\theta} \mathcal{L}_{diffuse}+\lambda \mathcal{L}_{deform},
    \label{eq:loss}
\end{equation}
where $G_\theta$ is the network of DDM with learnable parameters $\theta$, $\mathcal{L}_{diffuse}$ is the diffusion loss, $\mathcal{L}_{deform}$ is the deformation loss, and $\lambda$ is a hyper-parameter. 

Specifically, the diffusion loss is to estimate the latent code $c$, which is originally designed for the reverse diffusion process \cite{ho2020denoising}. Thus, for the source $S$ and the target $T$, the diffusion loss is calculated by:
\begin{equation}
    \mathcal{L}_{diffuse}=\mathbb{E}_{\epsilon, x, t}||G_{\theta}^{diffuse}(x_t, t; S, T) - \epsilon ||_2^2,
\end{equation}
where $G_{\theta}^{diffuse}$ is the diffusion module of DDM whose output is $c$, $\epsilon$ is the Gaussian distribution data with $\mathcal{N}(0, \textbf{I})$, and $t$ is a uniformly sampled time step in $[0, u]$. On the other hand, the deformation loss is to generate the deformation fields to warp the source image into the target. Based on the energy function of classical image registration, we design the deformation loss as follows:
\begin{equation}
    \mathcal{L}_{deform}=-{NCC}(S(\phi), T)+\lambda_R \Sigma ||\nabla \phi||^2,
    \label{eq:loss_deform}
\end{equation}
where $\phi$ is the output of the deformation module $G_{\theta}^{deform}$ using the input of the latent code $c$ and the source image $S$, $NCC$ is the local normalized cross-correlation \cite{balakrishnan2018unsupervised}, and $\lambda_R$ is a hyper-parameter. We set $\lambda_R=1$ in our experiments. In (\ref{eq:loss_deform}), the first term is the dissimilarity metric between the deformed source image and the target, and the second term is the smoothness regularization for the deformation fields.

Accordingly, using the loss function (\ref{eq:loss}), our model is trained by end-to-end learning in an unsupervised manner. This allows the diffusion module to learn the score function of the deformation by comparing the source and target. That is, the latent code $c$ from the diffusion module can provide spatial information between the source and target, which enables the deformation module to generate the deformation fields of the source toward the target image.

\subsection{Intermediate Frames for Temporal Volume Generation}
Once the proposed DDM is trained, for a given condition of $S$ and $T$, our model produces a deformed source image aligned with the target when $x_0$ is set to the target $T$. Here, when the latent code passed to the deformation module is set to zero, our model produces deformation fields that hardly deform the source image. Thus, the 4D temporal images from the source to the target can be obtained by adjusting the latent code $c$. 

\begin{figure}[t!]
\centering
\includegraphics[width=\textwidth]{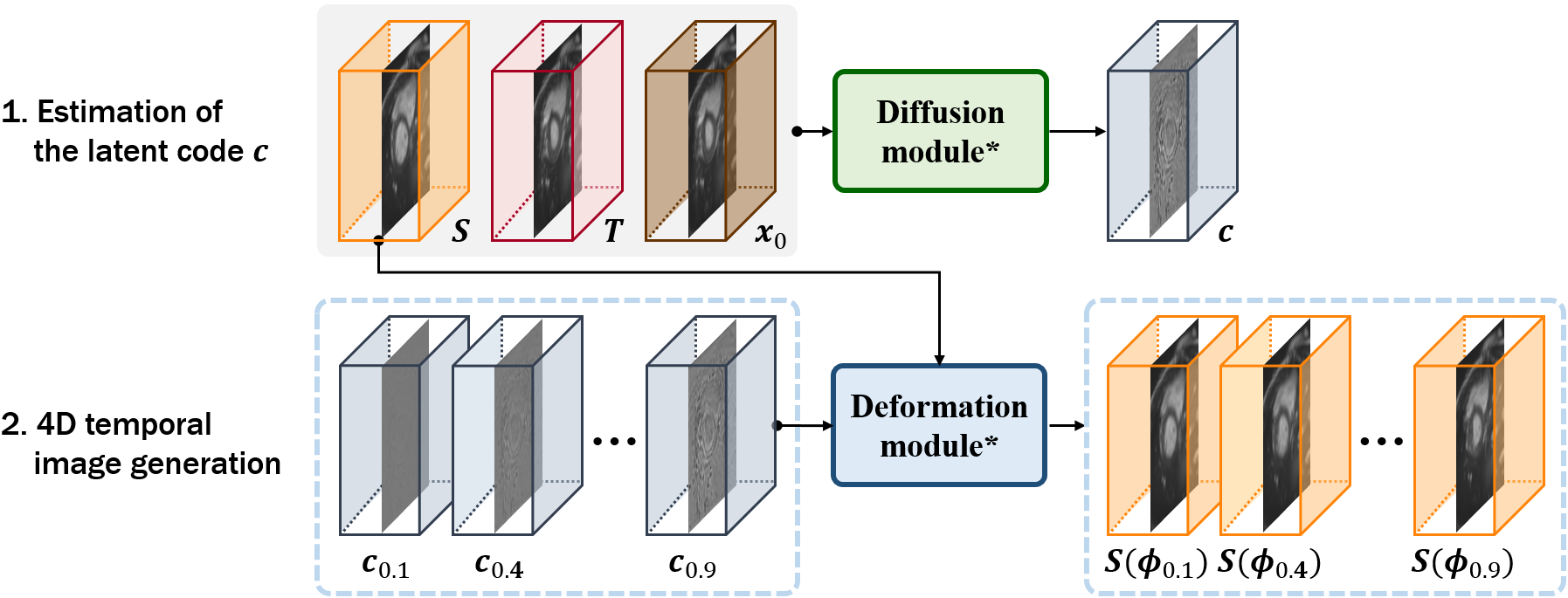}
\caption{The proposed inference flow for 4D temporal image generation. Once the latent code $c$ is estimated, the continuous deformed frames between the source and the target are generated by scaling the latent code $c$ with $\gamma$, i.e. $c_\gamma$. } 
\label{fig:inference}
\end{figure}

Specifically, as shown in Fig.~\ref{fig:inference}, the latent code $c$ is first estimated by the diffusion module with the fixed parameters $\theta^\ast$:
\begin{equation}
    c = G_{\theta^\ast}^{diffuse}(S, T, x_0),
\end{equation}
where $x_0$ is the target $T$. Then, since this latent code provides the score function of deformation between the source and the target, the intermediate frame is generated by warping the source $S$ with the deformation fields $\phi_\gamma$:
\begin{equation}
    \phi_\gamma =  G_{\theta^\ast}^{deform}(S, c_\gamma),
\end{equation}
where $c_\gamma$ is the latent code adjusted by scaling $c$ by $\gamma\in [0, 1]$, i.e. $c_\gamma=\gamma \cdot c$. Therefore, through the simple interpolation of the latent code, our model can generate temporally continuous deformed images along the trajectory from the source to the target image.

\section{Experiments}
\subsubsection{Dataset and Metric}
To verify the proposed method for 4D image generation, we used the publicly available ACDC dataset \cite{bernard2018deep} that contains 100 4D cardiac MRI data.  We trained and tested our model to provide the 4D temporal images from the end-diastolic to the end-systolic phases. All MRI scans were re-sampled with a voxel spacing of $1.5\times 1.5\times 3.15 mm$ and cropped to $128\times 128 \times 32$. The image intensity is normalized to [-1, 1]. The dataset was split into 90 and 10 scans for training and testing. 

To quantify the image quality, we used the peak signal-to-noise ratio (PSNR) and the normalized mean square error (NMSE) between the generated deformed images and the real temporal data. Also, since the dataset provides the manual segmentation maps on several cardiac structures of the end-diastolic and systolic volumes for each subject, to evaluate the accuracy of the diastolic-to-systolic deformation, we computed the Dice score between the ground truth annotations and the estimated deformed maps.

\begin{figure}[b!]
\centering
\includegraphics[width=\textwidth]{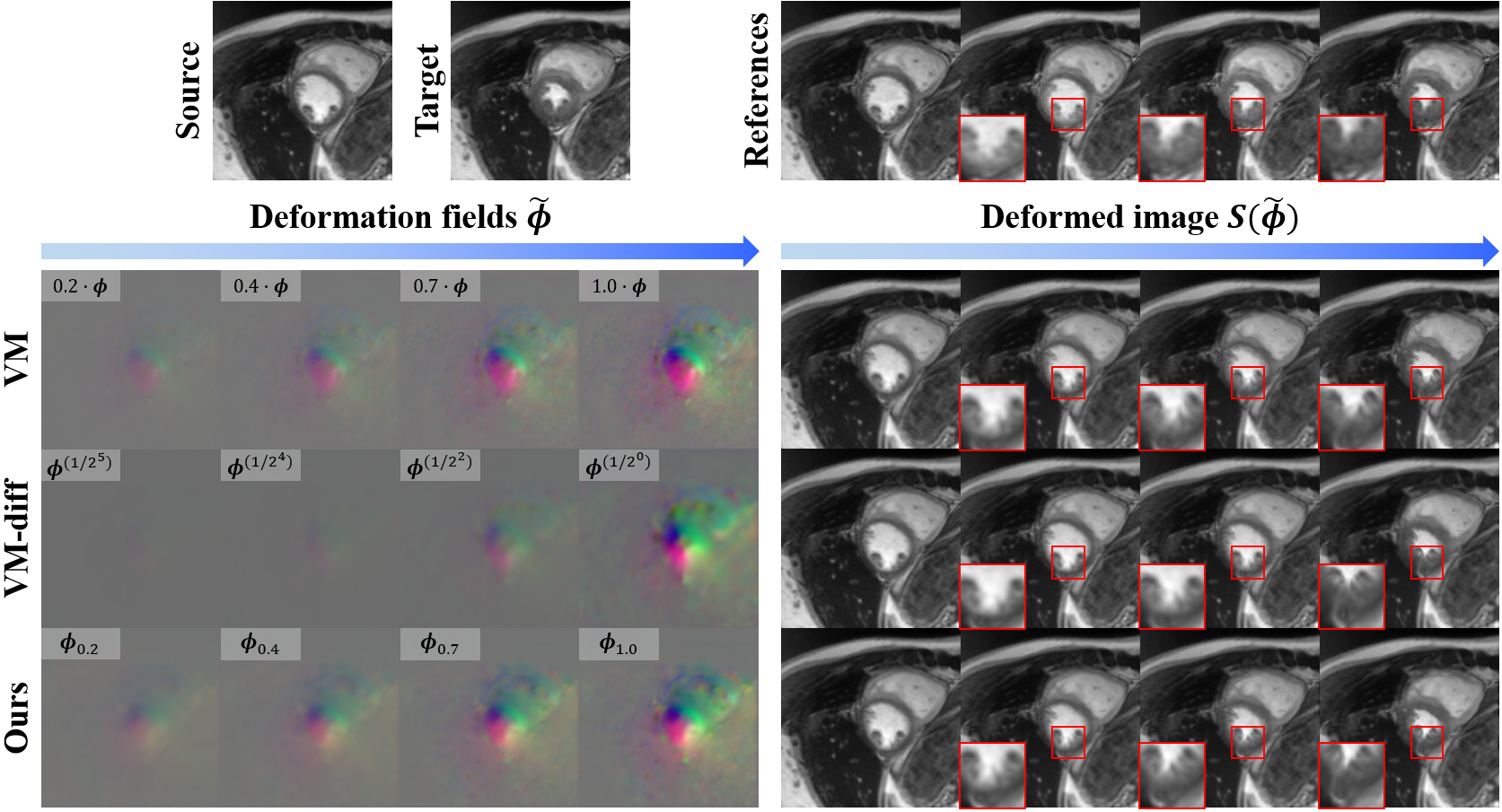}
\caption{Visual comparison results of the temporal cardiac image generation. Using the source and target, the deformed intermediate frames $S(\tilde{\phi})$ (right) are generated using the deformation fields $\tilde{\phi}$ (left). The references are from the ground-truth 4D images.} 
\label{fig:result_qual}
\end{figure}

\subsubsection{Implementation Details}
We built our proposed method on the 3D UNet-like structures that has the encoder and decoder with skip connections. For the diffusion module, we employed the network designed in DDPM \cite{ho2020denoising} and set 8, 16, 32, and 32 channels for each stage. We set the noise level from $10^{-6}$ to $10^{-2}$, which was linearly scheduled with $u=2000$. For the deformation module, we used the architecture of VoxelMorph-1 \cite{balakrishnan2018unsupervised}. We adopt Adam algorithm \cite{kingma2014adam} with a learning rate $2\times 10^{-4}$, and trained the model with $\lambda=20$ for 800 epochs by setting the batch size as 1. The proposed method was implemented with PyTorch 1.4 \cite{paszke2017automatic} using an Nvidia Quadro RTX 6000 GPU. The memory usage for training and testing was about 3GB and 1GB, respectively. Also, the model training takes 6 hours, and the testing takes an average of 0.456 seconds. The source code is available at \url{https://github.com/torchDDM/DDM}.

\subsubsection{Results and Discussion}
We compared ours with the learning-based registration models: VM \cite{balakrishnan2018unsupervised} and VM-diff \cite{dalca2018unsupervised}. The VM can provide the intermediate frames between the source and target by scaling the estimated deformation fields, i.e. $\gamma \cdot \phi$, while the VM-diff with diffeomorphic constraint can give the deformed images by integrating the velocity field with the timescales, i.e. $\partial \phi^{(t)} / \partial t = v(\phi^{(t)})$. We implemented these methods using the source code provided by the authors with the recommended parameters. For the VM, the weight of smooth regularization was set to 1. For the VM-diff, the number of integration steps, the noise parameter, and the prior lambda regularization for KL loss were set to 7, 0.01, and 25, respectively. For a fair comparison, we used the same network with our deformation module and trained with the learning rate $2\times 10^{-4}$ until the models converge. At the inference, we set the number of deformations according to the number of frames in each ground-truth 4D data.

\begin{table}[t!]
\caption{Quantitative comparison results of the average PSNR, NMSE, Dice, and test runtime. Standard deviations are shown in parentheses. Asterisks denote statistical difference of the baseline methods over ours ($\ast\ast$: $p<0.005$ and $\ast$: $p<0.05$).}
\label{tab:result_quan}
\centering
\resizebox{\linewidth}{!}{
\begin{tabular}{M{0.9cm}|M{1.9cm}|M{2.6cm}|M{2.6cm}|M{2.55cm}|M{1.9cm}}
\hline
Data & Method &  PSNR (dB) $\uparrow$ & NMSE ($\times 10^{-8}$) $\downarrow$ & Dice  $\uparrow$ & Time (sec) $\downarrow$ \\
\hline
\multirow{2}{*}{Train} & \multicolumn{1}{l|}{Initial} &  \multicolumn{1}{l|}{\,29.683 (3.116)} & \multicolumn{1}{l|}{\;\,0.690 (0.622)} & \multicolumn{1}{l|}{\;\;0.700 (0.185)} & - \\
& \multicolumn{1}{l|}{DDM (Ours)} & \multicolumn{1}{l|}{\,32.788 (2.859)} & \multicolumn{1}{l|}{\;\,0.354 (0.422)} & \multicolumn{1}{l|}{\;\;0.830 (0.112)} & 0.456 \\
\hline
\multirow{4}{*}{Test} & \multicolumn{1}{l|}{Initial} &  \multicolumn{1}{l|}{\,28.058 (2.205)} & \multicolumn{1}{l|}{\;\,0.790 (0.516)} & \multicolumn{1}{l|}{\;\;0.642 (0.188)} & - \\
& \multicolumn{1}{l|}{VM}      &  \multicolumn{1}{l|}{\,30.562 (2.649) $\ast$} & \multicolumn{1}{l|}{\;\,0.490 (0.467) $\ast$} & \multicolumn{1}{l|}{\;\;0.784 (0.116) $\ast$} & 0.219 \\
& \multicolumn{1}{l|}{VM-diff} &  \multicolumn{1}{l|}{\,29.481 (2.473) $\ast\ast$} & \multicolumn{1}{l|}{\;\,0.602 (0.477) $\ast\ast$} & \multicolumn{1}{l|}{\;\;0.794 (0.104)} & 2.902 \\
& \multicolumn{1}{l|}{DDM (Ours)} & \multicolumn{1}{l|}{\,\textbf{30.725} (2.579)} & \multicolumn{1}{l|}{\;\,\textbf{0.466} (0.432)} & \multicolumn{1}{l|}{\;\;\textbf{0.802} (0.109)} & 0.456 \\
\hline
\end{tabular}}
\end{table}

\begin{figure}[t!]
\centering
\includegraphics[width=\textwidth]{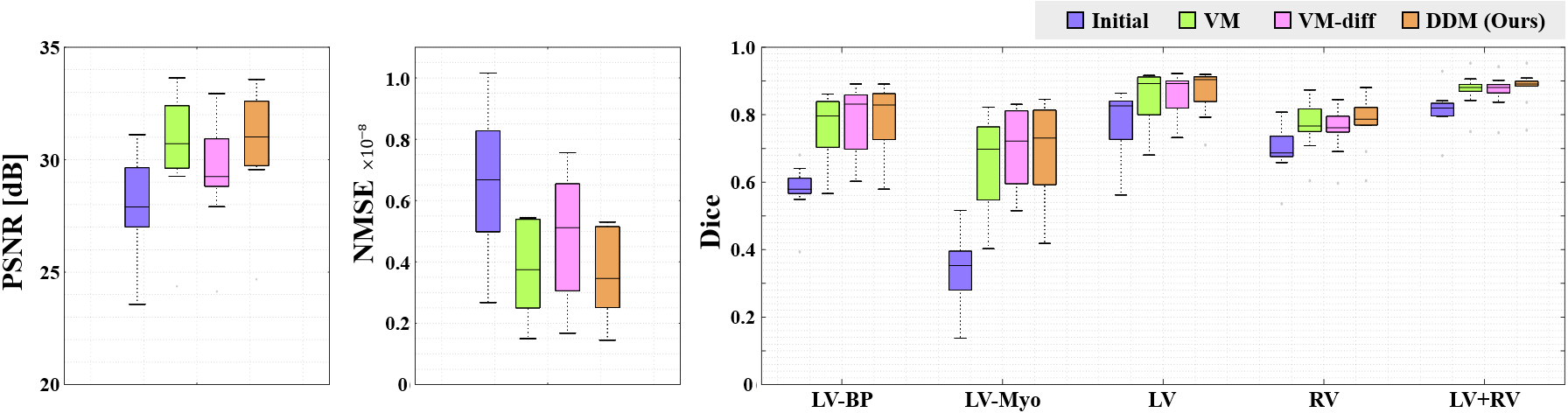}
\caption{Boxplots of quantitative evaluation results on test data. PSNR and NMSE are computed using the generated frames from the diastolic to the systolic phases. Dice score is computed for the segmentation maps of the cardiac structures at the end-systolic phase: left blood pool (LV-BP), myocardium (LV-Myo), epicardium of the left ventricle (LV), right ventricle (RV), and the total cardiac region (LV+RV).} 
\label{fig:result_comp}
\end{figure}

Fig.~\ref{fig:result_qual} shows visual comparisons of the temporal image generation along the trajectory between the source and the target. We can observe that the deformation fields of VM only vary in scale, but their relative spatial distributions do not change, resulting in the scaled movement of anatomical structures. Also, the VM-diff with the integration of the velocity fields hardly deforms the source in the beginning, but sharply in the end. On the other hand, in the proposed method, the deformation fields estimated from the scaled latent codes represent the dynamic changes depending on the positions. Accordingly, the generated intermediate deformed images have distinct changes from the source to the target. 

The quantitative evaluation results are reported in Table~\ref{tab:result_quan} and Fig.~\ref{fig:result_comp}. Specifically, compared to the VM and VM-diff, our proposed DDM outperforms with higher PSNR and lower NMSE in generating continuous deformations from the diastolic to the systolic phase volumes. The average Dice score for the segmentation maps of cardiac structures also shows that ours achieves 80.2\% with about 1\% gain over the baseline methods. We can observe that the proposed method tested on training data shows similar gains for all metrics when compared to the results on test data.
Moreover, we evaluated the statistical significance of the results and observed that our model outperforms the VM on all metrics with p-values<0.05 under both paired t-test and Wilcoxon signed rank test. It is also worth noting that our model shows slightly higher accuracy on Dice score over the VM-diff with no significant difference but achieves significant improvement on PSNR and NMSE with p-value<0.001 under paired t-test and p-value<0.005 under Wilcoxon signed rank test. In addition, the average test time of VM-diff takes 2.902 seconds, while that of our DDM takes 0.456 seconds. These indicate that our method shows superiority in generating continuous deformations.

\begin{figure}[t!]
\centering
\includegraphics[width=\textwidth]{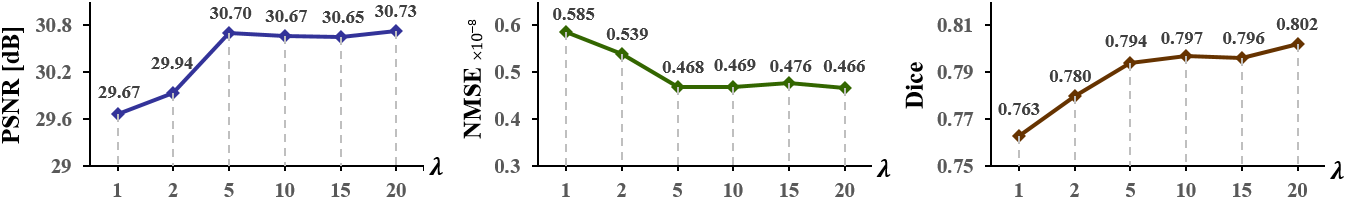}
\caption{Quantitative results over various hyper-parameter $\lambda$ in the proposed loss function. The graphs show the average values of PSNR, NMSE, and Dice metrics. } 
\label{fig:result_hyper}
\end{figure}


Furthermore, to analyze the effect of our designed loss function (\ref{eq:loss}), we studied our method with various hyper-parameter $\lambda$ values. Fig.~\ref{fig:result_hyper} shows the results of PSNR, NMSE, and Dice metrics. When $\lambda$ increases, PSNR increases and NMSE decreases initially, and they converge when it exceeds a certain level ($\lambda=5$), suggesting that the large value of deformation loss does not affect the generation performance more. In contrast, the Dice score improves according to the $\lambda$ increases, which suggests that the deformation loss affects the registration accuracy, and also helps the diffusion module to estimate latent codes to generate temporally continuous deformed images along the trajectory. On the other hand, when we trained our model only using the deformation loss, PSNR, NMSE, and Dice scores were 30.72, 0.473$\times 10^{-8}$, and 0.799, respectively, which were lower than the optimal results of our method. This indicates that the diffusion loss is effective to learn the latent code for generating temporal images.

\section{Conclusion}
We propose a novel 4D image generation framework by adapting the denoising diffusion probabilistic model to the deformable registration model. Our method learns the distribution of the source and target and estimates the latent code to generate deformed images along the continuous trajectory. Experimental results on 4D cardiac MR image generation verify that the proposed method produces dynamic deformations from the end-diastolic to systolic phase volumes, and outperforms the existing registration-based models. This makes our model a promising tool in clinical applications as it provides intermediate frames between two images in real-time when analyzing changes in anatomical structures.

\subsubsection{Acknowledgements} 
This work was supported by Institute of Information \& communications Technology Planning \& Evaluation (IITP) grant funded by the Korea government(MSIT) (No.2019-0-00075, Artificial Intelligence Graduate School Program(KAIST)), by the National Research Foundation of Korea under Grant NRF-2020R1A2B5B03001980, by the Korea Medical Device Development Fund grant funded by the Korea government (the Ministry of Science and ICT, the Ministry of Trade, Industry and Energy, the Ministry of Health \& Welfare, the Ministry of Food and Drug Safety) (Project Number: 1711137899, KMDF\_PR\_20200901\_0015), and by the MSIT(Ministry of Science and ICT), Korea, under the ITRC(Information Technology Research Center) support program(IITP-2021-2020-0-01461) supervised by the IITP(Institute for Information \& communications Technology Planning \& Evaluation).


%
%
%

%
\end{document}